\documentstyle[12pt, psfig]{article}
\def\f{\frac}
\begin{document} 

\title{Transverse Energy production\\
in Ultrarelativistic Heavy Ion Collisions
\footnote{This work was supported by 
the U.S. Department of Energy under contract No. DE-FG02-93ER40764,
DE-FG-02-92-ER40699, and DE-AC02-76CH00016}}
\author{Bin Zhang$^1$, Yang Pang$^{1, 2}$, and Miklos Gyulassy$^1$\\
{\em $^1$Physics Department, Columbia University,}\\
{\em New York, NY 10027, USA}\\
{\em $^2$Brookhaven National Laboratory,}\\
{\em Upton, New York, 11973}
}
\date{}
\maketitle

\begin{abstract}
Kinetic theory is used to check the applicability of parton cascade
in 1 dimensional expansion. Using the information provided by 3 dimensional
parton cascade, we model the transverse expansion by an effective area. With
this model, kinetic theory is able to give prediction of the time development
of transverse energy which is in good agreement with the parton cascade
results.
\end{abstract}

\newpage


\section{Introduction}

Cascade method has long been used to study nucleus-nucleus
collisions \cite{cascade1}. Recently, the OSCAR standard \cite{oscar1}has been
passed to provide some objective scientific criteria for cascade
simulations. One goal of the open standard working group is to develop and
perform a set of standardized tests to ensure the applicability of the cascade
method. One test we proposed is to compare the cascade prediction for time
dependence of transverse energy with the scaling kinetic theory
prediction. It's shown that ZPC gives result which is in good agreement with
the kinetic theory prediction when the isolated 2 body collision criteria is
satisfied. The we go from 1 dimensional expansion to 3 dimensional
expansion. The cascade gives us the transverse expansion information. With a
model of the transverse expansion, we see the kinetic theory gives consistent
result with the cascade prediction.

\section{Boost Invariant Relativistic Transport}

Parton transport equation:
\[(p\cdot\partial_x)f(x, p) = (p\cdot u)(C(x, p) + S(x, p))\]
can be simplified by changing variables to: $\tau$, $\xi = \eta - y$,
and $p_T$ \cite{kajantie1}.

We put the partons in on a hyperbola of constant $\tau_0$ with Boltzmann
distribution, i.e., the source term is:
\[S(\tau,\xi,p_T) = C \delta(\tau-\tau_0) exp(-\f{p_T}{T_0}cosh\bar{\xi}).
\]

Under relaxation time approximation, the solution of the above transport
equation is:
\[f(\tau,\xi,p_T)=\Theta(\tau-\tau_0)exp(-\int_{\tau_0}^\tau
\f{d\tau'}{\tau_c(\tau')})\;C\;exp(-\f{p_T}{T_0}cosh\xi(\tau_0))\]
\[+\int_{\tau_0}^\tau\;d\tau'
\f{exp(-\int_{\tau_0}^\tau\f{d\tau"}{\tau_c(\tau")})}{\tau_c(\tau')}
f_{eq}(\tau',\xi'(\tau'),p_T)\]

The proper energy density
\[\epsilon(\tau)=\int\;d\xi d^2p_T(p_Tcosh\xi)^2
f(\tau,\xi,p_T)\]
and the transverse energy per unit rapidity 
\[e_T(\tau) = 
\f{dE_T}{dy}=\tau A \int d\eta d^2p_T p_T^2 cosh\xi f(\tau,\xi,p_T)\]
can be calculated from the above phase space distribution. We get:
\[\epsilon(\tau)=\Theta(\tau-\tau_0)exp(-\int_{\tau_0}^\tau
\f{d\tau'}{\tau_c(\tau')})\f{\tau_0}{\tau}h(\f{\tau_0}{\tau})\epsilon(\tau_0)\]
\[+\int_{\tau_0}^\tau\;d\tau'
\f{exp(-\int_{\tau_0}^\tau\f{d\tau"}{\tau_c(\tau")})}{\tau_c(\tau')}
\f{\tau'}{\tau}h(\f{\tau'_0}{\tau})\epsilon(\tau'),\]
in which:
\[h(x)=\f{1}{2}(|x|+\f{Arcsin\sqrt{1-x^2}}{\sqrt{1-x^2}}),\]
and
\[e_T(\tau)=\Theta(\tau-\tau_0)exp(-\int_{\tau_0}^\tau
\f{d\tau'}{\tau_c(\tau')})e_T(\tau_0)\]
\[+\f{\pi}{4}A\int_{\tau_0}^\tau\;d\tau'
\f{exp(-\int_{\tau_0}^\tau\f{d\tau"}{\tau_c(\tau")})}{\tau_c(\tau')}
\tau'\epsilon(\tau')\]

The relaxation time
\[\tau_c=\f{1}{n\sigma_\theta},\]
in which 
\[\sigma_\theta=\int sin^2\theta \f{d\sigma}{d\Omega}d\Omega\]
is the transport cross section.

We take the differential cross section to be:
\[\f{d\sigma}{d\hat{t}}=\f{9 \pi \alpha^2}{2}(\f{\mu^2}{\hat{s}}+1)
/(\hat{t}-\mu^2)^2,\]
such that the total cross section is
\[\sigma = \f{9 \pi \alpha^2}{2 \mu^2}\]

Notice many authors have studied the energy density equation \cite{kajantie1}
\cite{baym1}, and a
variable similar to transverse energy (proper time times the energy density) is
introduced for calculational convenience. 
But experiments measure only transverse energy, and the
transverse energy is not simply a product of proper time, transverse area and
the energy density. It can not be simply calculated from one integral equation,
but has to be calculated from a set of couple integral equations including the
energy density evolution equation.

The parton cascade gives reasonably good result for the time development of 
the transverse energy per unit rapidity as seen in Fig. 1. 

\section{Effect of transverse expansion on the transverse energy production}

It is difficult to solve kinetic equation semianalytically with the transverse
expansion. But we can get the transverse expansion information from the cascade
and using a model for the transverse expansion to make the semianalytical
calculation possible.

A model for freezeout is to assume that the $e_T$ freezes out outside the
rarefaction front, i.e.,
\[\f{de_{T{\em out}}(\tau)}{d\tau} = 
- \f{e_{T{\em in}}}{A(\tau)}\f{dA(\tau)}{d\tau}\]
The contribution to $e_T$ from inside the rarefaction front is just the
rescaled (by the effective area inside the rarefaction front) contribution
derived before:
\[e_{T{\em in}}(\tau)=\f{A(\tau)}{A(\tau_0)}
\Theta(\tau-\tau_0)exp(-\int_{\tau_0}^\tau
\f{d\tau'}{\tau_c(\tau')})e_T(\tau_0)\]
\[+\f{\pi}{4}A(\tau)\int_{\tau_0}^\tau\;d\tau'
\f{exp(-\int_{\tau_0}^\tau\f{d\tau"}{\tau_c(\tau")})}{\tau_c(\tau')}
\tau'\epsilon(\tau').\]


The cascade gives the weight function for the transverse area:
\[w(r,\tau)=\left\{
\begin{array}{ll}
1 & r<R-v_c\tau \\
(\f{R+v_c\tau-r}{2v_c\tau})^4 & R-v_c\tau<r<R+v_c\tau \\
0 & r>R+v_c\tau
\end{array}
\right.
\]

This weight function is almost independent of interaction.
The effective transverse area can be calculated from:
\[A(\tau)=\int_0^\infty w(r,\tau)dr\]

The model results (Fig. 2) are similar to the cascade prediction. Both of them
give more transverse energy production than the 1-d expansion.

\section{Conclusion}

For the 1 dimensional expansion, ZPC's prediction for the
time development of the transverse energy production is consistent with the
kinetic theory prediction. With the model motivated by the cascade, we can
perform the semianalytic calculation of the kinetic theory and give out
consistent result with the cascade's.


{}

\newpage
\begin{figure}[h]
\vspace{2.5cm}
\hspace{1.01cm}
\psfig{figure=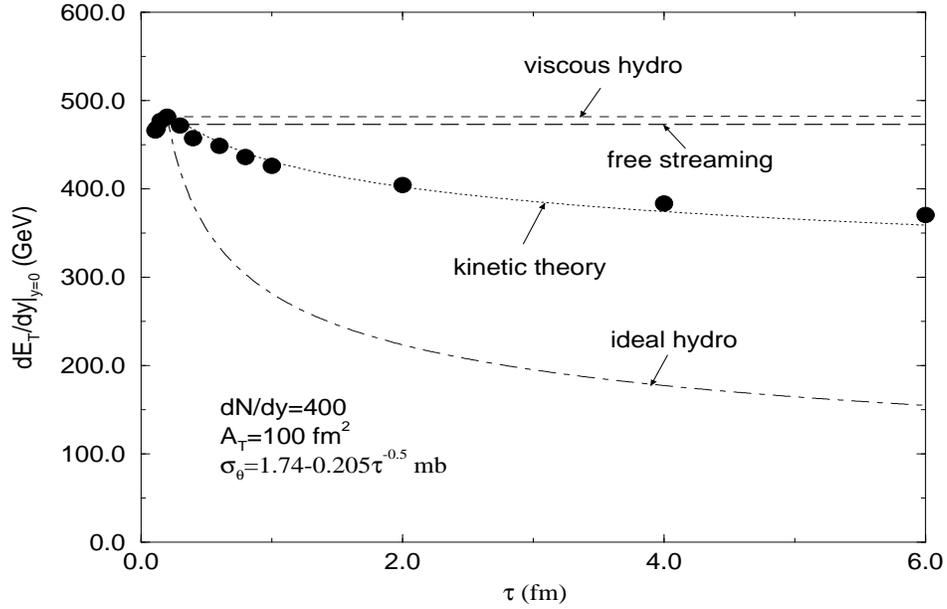,height=3.5in,width=3.0in,angle=-90} 
\vspace{-0.5cm}
\caption{
  20 event averaged 1-d expansion. Initially (at $\tau=0.1\;fm$),
  $T_0=500\;MeV$, $\eta\in[-5,5]$,
  $\f{dN}{d\eta}=400$. Screen mass $\mu = 3\;fm^{-1}$, strong interaction
  coupling constant $\alpha_S=0.47$. Interaction length $0.3\;fm$. Initial
  mean free path $0.1\;fm$, smaller than the interaction length. This is the
  reason that initially the cascade gives a slight increase. The comparison
  starts at $\tau=0.2\;fm$ when the mean free path is close to the interaction
  length and good agreement is achieved.
}
\end{figure}

\newpage
\begin{figure}[h]
\vspace{2.5cm}
\hspace{1.01cm}
\psfig{figure=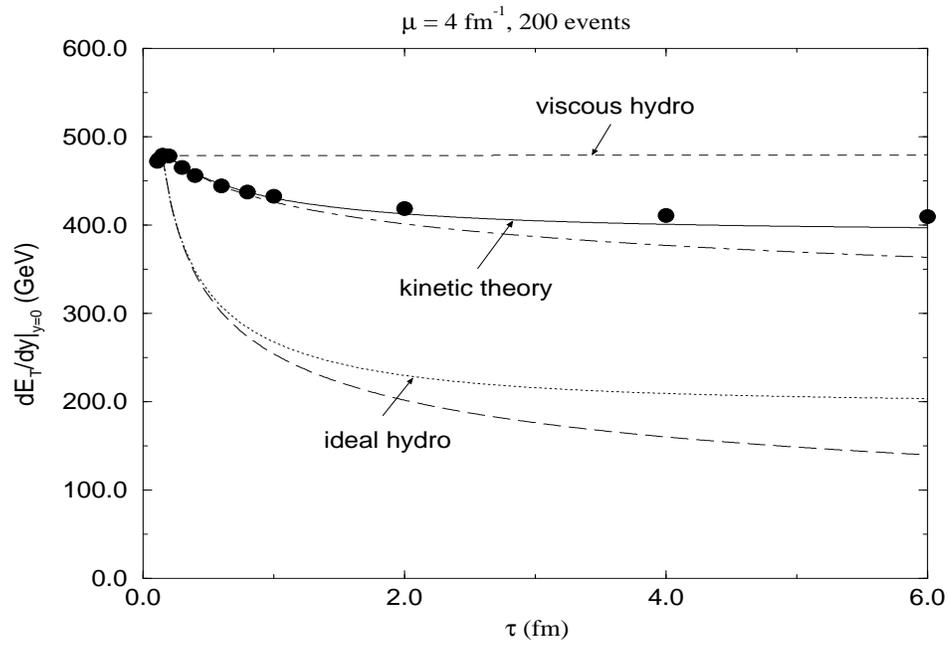,height=3.5in,width=3.0in,angle=-90} 
\vspace{-0.5cm}
\caption{
  Comparison of 3 dimensional expansion cascade and kinetic theory results. The
  initial condition is similar to 1-d case except that the transverse positions
  of particles are within a disc of radius $5\;fm$. The dash-dotted and the
  long dashed curve are corresponding 1-d kinetic theory and ideal hydro
  predictions for comparison.
}
\end{figure}

\end{document}